\documentclass[aps,prl,superscriptaddress,twocolumn]{revtex4}
\usepackage{amsmath}
\usepackage{amssymb}
\usepackage{amsthm}
\usepackage{graphicx}
\usepackage{graphics}
\usepackage{subfigure}
\usepackage{bm}
\usepackage{hyperref}
\usepackage{rotating}

\def\beas{\begin{eqnarray*}}
\def\eeas{\end{eqnarray*}}
\def\bea{\begin{eqnarray}}
\def\eea{\end{eqnarray}}
\def\be{\begin{equation}}
\def\ee{\end{equation}}

\newcommand{\down}{\downarrow}

\newcommand{\si}{\sigma}

\newcommand{\bpm}{\begin{pmatrix}}
\newcommand{\epm}{\end{pmatrix}}
\newcommand{\bmm}{\begin{matrix}}
\newcommand{\emm}{\end{matrix}}
\newcommand{\up}{\uparrow}

\begin{document}

\title{Fractional topological insulators}

\author{Michael Levin}
\affiliation{Department of Physics, Harvard University, Cambridge,
Massachusetts, 02138, USA}
\affiliation{Department of Physics, University of California, Santa
Barbara, 93109, USA}
\author{Ady Stern}
\affiliation{Department of Condensed Matter Physics, Weizmann
Institute of Science, Rehovot 76100, Israel}

\date{\today}

\begin{abstract}
We analyze generalizations of two dimensional topological
insulators which can be realized in interacting, time reversal invariant
electron systems. These states, which we call fractional
topological insulators, contain excitations with
fractional charge and statistics in addition to protected edge modes.
In the case of $s^z$ conserving toy models, we show that a system is a
fractional topological insulator if and only if $\sigma_{sH}/e^*$
is odd, where $\sigma_{sH}$ is the spin-Hall conductance in units
of $e/2\pi$, and $e^*$ is the elementary charge in units of $e$.
\end{abstract}

%\pacs{}

\maketitle

% ----------------------------------------------------------------

\textsl{Introduction:} 
Recently, it was realized that in two 
dimensions there are two distinct universality classes of time 
reversal invariant band insulators - topological insulators and 
trivial insulators. \cite{kanemele,BZ0602}
The two kinds of insulators can be distinguished by
the fact that the edge of a topological insulator contains a
protected pair of gapless edge modes of opposite chiralities,
while no such protected edge modes exist for a trivial insulator.
Indeed, such a protected edge mode has been observed in $HgTe$
quantum wells.
\cite{KWB0766}

In interacting electron systems, many other gapped, charge-conserving
electronic ground states are possible in addition to band insulators. It
is natural to wonder: do analogues of topological and trivial 
insulators exist for these more general states? That is, do some of 
these states have time reversal protected edge modes, while some do 
not? This is the issue we investigate in this paper.

Our starting point is the standard toy model construction of
topological insulators. \cite{kanemele, BZ0602} In this
construction, one imagines a two-dimensional system of electrons in the
continuum where the electrons experience a spin-dependent magnetic field
$B_0{\hat z}\sigma_z$. (In a more realistic context, this kind of
physics can originate from spin-orbit coupling). \cite{BZ0602,kanemele}
One then assumes that the electrons are non-interacting and that
the electron density is tuned to an integer Landau filling $\nu =
k$. The ground state is thus made up of two decoupled spin species
which form integer quantum Hall states with opposite chiralities.
One can show that when $k$ is odd, this state is a topological
insulator; when $k$ is even it is a trivial insulator.

A simple way to generalize this construction is to imagine that
the two spin species each form \emph{fractional} quantum Hall (FQH)
states. Such states can be
realized in a toy model similar to the one described above. The
only new element is that one introduces a toy electron-electron
interaction where electrons of the same spin interact via a
short-range two-body repulsive force, and
electrons of different spin do not interact at all. This system
can then be mapped onto two decoupled FQH systems of the same
filling factor and opposite magnetic field. \cite{FNS0428} 
Depending on the electron density and the details of the electron 
interaction, one can engineer scenarios where each spin species 
forms arbitrary FQH states. 

In this paper, we study the properties
of these states and their stability to perturbations. We leave
the analysis of their experimental feasibility to future work.
Such states were considered by Bernevig et al. in the
context of the fractional quantum spin Hall effect. \cite{BZ0602}
Here, we show that these states are interacting analogues of
topological/trivial insulators: some of these states,
which we dub ``fractional topological insulators", have time
reversal protected edge modes and some, which we dub ``fractional
trivial insulators", do not. We find that a state is a fractional
topological insulator if and only if the integer 
parameter $\si_{sH}/e^*$ is odd, where $\si_{sH}$ is the spin-Hall 
conductance measured in units of $e/2\pi$, and 
$e^*$ is the elementary charge (e.g. smallest charge 
of any quasiparticle excitation), measured in units of $e$. 
This result is quite general and holds for any $s^z$ conserving model.

We also show that every fractional trivial
insulator with $1/e^*$ odd has a ``partner" fractional topological
insulator which is adiabatically equivalent to it in the absence
of time reversal symmetry. On the other hand, when $1/e^*$ is even, we
find that fractional topological insulators are not possible at
all (at least for the models described above). In the course of analyzing
this case, we derive a result about general electronic FQH states
which may be useful in its own right: we show that if $1/e^*$ is
even then $\sigma_{xy}/e^*$ must also be even. One implication is
that if $\sigma_{xy} = 1/2$ then $e^*$ is at most $1/4$. Note that
there are no such restrictions on the charge of the \emph{lowest
energy} excitation; in principle this can be any multiple of $e^*$.

An interesting example where our results are applicable
occurs in the case where the two spin species each form $\nu = 1/2$
FQH states. More specifically, consider the six
possibilities corresponding to the Pfaffian state \cite{MR9162},
strong-pairing state \cite{H8375}, $331$ state \cite{H8375,HR8856},
and their particle-hole conjugates. These $6$
states all have different edge structures - some states have one
chiral boson mode (strong-pairing), some have two ($331$ and
anti-strong-pairing), some have three (anti-$331$),
and some have Majorana modes (Pfaffian and anti-Pfaffian). \cite{W9355}
One might guess that some of these states are fractional
topological insulators - for example those with an odd number of
chiral boson edge modes. However, this is \emph{not} the case:
these states all have $\sigma_{sH} = 1/2$, $e^* = 1/4$, so that
$\sigma_{sH}/e^*$ is even. In fact, this is
true for any $\nu = 1/2$ state.

\textsl{Flux insertion argument:}
We derive the above criterion using a generalization of the flux insertion 
argument of Fu et al. \cite{FK0612} We first review this argument in the
case of the above non-interacting toy model. Consider
a cylindrical geometry, and assume that the number of
electrons is even. In addition, assume that the ground state is
time reversal invariant when there is zero flux through the cylinder.
Under these assumptions, Fu et al. argued that there must be at least
one low-lying excited state if $k$ is odd - even in the presence
of an arbitrary time reversal invariant perturbation. To see
this, start with the  ground state of the toy model at zero flux and
imagine adiabatically inserting half of a flux quantum
$\Phi_0/2$ through the cylinder. Let us call the resulting state
$\Psi_1$. Similarly, let $\Psi_2$ be the state obtained by
adiabatically inserting $-\Phi_0/2$ flux through the
cylinder. The state $\Psi_2$ can be transformed into $\Psi_1$ by
inserting a full flux quantum - an operation which
transfers $k$ electrons with spin up from the left edge to the
right edge, and $k$ electrons with spin down from the right edge
to the left edge. In particular, this means that the two states are
orthogonal. On the other hand, $\Psi_1, \Psi_2$ have the same energy
since they are time reversed partners: $\Psi_2 = \mathcal{T}\Psi_1$.
Thus, the system with half of a flux quantum has a degenerate,
low-lying eigenstate. The key point is that this degeneracy is
robust against arbitrary time reversal invariant perturbations
if $k$ is odd. The reason is that when $k$ is odd, there are an
odd number of unpaired electrons localized near each of the two
edges of $\Psi_1,\Psi_2$. Thus, as long as the edges are well
separated, Kramer's theorem guarantees that $\Psi_1, \Psi_2$
are degenerate (and in fact there must be two other degenerate
states, in addition). The claim now follows:
the robust degeneracy at half of a flux quantum implies that
there is also a robust low-lying excited state at zero flux
(since the insertion of half of a flux quantum cannot close a gap).

We now generalize this argument to the interacting toy models
described above (or more generally, any $s^z$ and charge conserving
system). The crucial difference with the non-interacting case
is that in the more general case the ground state may have
topological order. The presence of nontrivial topological order
means that in a cylindrical geometry there are always at least a
finite number of low-lying states - even if the edge is gapped.
\cite{degcyl} These low-lying states belong to different
topological sectors and can be distinguished from one another by
measuring the Berry phase associated with moving a quasiparticle
around the cylinder. Thus, to show that the edge is gapless in the
general case, it is not enough to just establish that there are
low-lying states at zero flux; we have to show that there are
low-lying states in the same topological sector as the ground
state.

Because we need to establish this stronger claim, the
generalized flux insertion argument begins by inserting not
$\pm \Phi_0/2$ flux but $\pm N \Phi_0/2$ flux
where $N$ is the smallest integer such that the resulting
states $\Psi_1, \Psi_2$ are in the same topological sector. The
argument then proceeds as before. One notes that $\Psi_1$ can be
obtained from $\Psi_2$ by transferring $N \sigma_{sH}$ spin up
electrons from the left edge to the right edge and $N \sigma_{sH}$
spin down electrons from the right edge to the left edge. If $N
\sigma_{sH}$ is odd, then there is a Kramers degeneracy associated
with each edge. In this case, the degeneracy between $\Psi_1,
\Psi_2$ (and the two other states) is stable and cannot be
split by any time reversal invariant perturbation. Since
$\Psi_1,\Psi_2$ are in the same topological sector, we conclude 
that there must be at least one low-lying state at zero flux in the same
sector as the ground state. Hence the edge cannot be gapped out
by any time reversal invariant perturbation.

To complete the argument, we need to determine the integer $N$.
To this end, note that $N$ can be equivalently defined as the
minimal number of flux quanta that need to be inserted to go from
an initial state to a final state in the same topological sector.
It is easy to see that adiabatically inserting $N$
flux quanta changes the Berry phase associated with braiding a
quasiparticle around the cylinder by
\begin{equation}
\Delta \theta = 2\pi N q
\label{berry}
\end{equation}
where $q$ is the charge of the quasiparticle (in units of $e$). In
order for the initial and final states to be in the same
topological sector, $\Delta \theta$ must be a multiple of $2\pi$
for every quasiparticle. Thus, the minimal value of $N$ is $1/e^*$,
with $e^*$ the charge of the smallest charged quasiparticle.
Using this value of $N$ in the above argument, we derive
the above criterion that there is a protected edge mode
whenever $\sigma_{sH}/e^*$ is odd.

\textsl{Microscopic analysis:}
While the flux insertion argument proves that there is a protected
edge mode whenever $\sigma_{sH}/e^*$ is odd, it does not prove
that the edge modes can be gapped out when $\sigma_{sH}/e^*$ is
even. In order to fill in this gap, and also to show that the two
degenerate states from the argument are part of a gapless
mode, we now rederive the criterion using a microscopic
approach. We focus on the abelian case for simplicity. In general,
the edge of an abelian FQH state is
described by a Lagrangian density \cite{W9505}
\begin{eqnarray}
L_c(\phi;K,V,t) &=& \frac{1}{4\pi} (K^{ij}\partial_x \phi_i
\partial_t \phi_j - V^{ij}\partial_x \phi_i \partial_x \phi_j
\nonumber \\
&+& \epsilon^{\mu\nu}t^i\partial_\mu\phi_i A_\nu)
\label{action-edge3}
\end{eqnarray}
Here, $\phi$ is an $N$-component vector of fields, $K$ is the
$N\times N$ K-matrix, $V$ is the velocity matrix, $t$ is the
charge vector, and $A_\mu$ is the external vector potential. We
use a normalization where
electron creation operators are of the form $e^{i\theta(l)}$ with
$\theta(l)\equiv l^TK\phi$ and $l$ an integer valued $N$
dimensional vector satisfying $l^Tt=1$.

For the time reversal symmetric systems that we study, there are
$N$ fields $\phi_i^\up, \phi_i^\down$ for each spin direction and
the Lagrangian density is of the form
\begin{equation}
L=L_c(\phi^\up;K,V,t)+L_c(\phi^\down;-K,V,t)
\label{achiral}
\end{equation}
The Lagrangian  (\ref{achiral}) has $2N$ gapless edge modes, $N$
for each chirality. Our goal is to find the conditions under
which these modes can be gapped out by charge conserving, time
reversal symmetric perturbations. The question we ask
is a question of principle, and therefore we will not discuss
whether particular terms exist in realistic conditions. We also
will not impose a requirement of momentum or spin conservation on
the terms we study, since both momentum and spin may be exchanged
with impurities or an underlying lattice.

The creation operators for electrons of the two spin directions
are $e^{i\theta^\up(l)}$ and $e^{-i\theta^\down(l)}$ with
$\theta^\up(l),\theta^\down(l)$ defined as above. We use the
convention that the fields transform under time reversal
as $\phi^\up \rightarrow  \phi^\down$, $\phi^\down \rightarrow
\phi^\up  - \pi K^{-1} t$. This guarantees that electron creation
operators transform as $\psi^\dagger_\up \rightarrow
\psi^\dagger_\down$, $\psi^\dagger_\down \rightarrow
-\psi^\dagger_\up$.

It is convenient for us to define a $2N$-dimensional vector $\Phi =
\bpm \phi^\up \\ \phi^\down \epm$, a $2N \times 2N$ K-matrix
$\mathcal{K}=\bpm K& 0 \\ 0 & -K\epm $ and a charge vector
$\tau = \bpm t \\ t \epm$. Also, let $\Sigma_x=\bpm 0 & \bf{1} \\
\bf{1} & 0\epm $ and $\Sigma_z=\bpm \bf{1} & 0 \\ 0 &-\bf{1} \epm $
where $\bf{1}$ is the $N$-dimensional unit matrix.
In this notation, a generic charge conserving scattering term is of
the form  $U(x)e^{i\Theta(\Lambda)}+h.c.$ where
$\Theta(\Lambda)\equiv \Lambda^T\mathcal{K}\Phi$, and $\Lambda$ is
a $2N$-dimensional integer valued vector satisfying $\Lambda^T
\tau = 0$. The field $\Theta$ transforms under time reversal
as
\begin{equation}
\mathcal{T}\Theta(\Lambda) \mathcal{T}^{-1}
=\Theta(-\Sigma_x\Lambda)-Q(\Lambda)\pi,
\label{timereversal}
\end{equation}
where $Q(\Lambda)\equiv \frac{1}{2}\Lambda^T\Sigma_z \tau$ is
the number of spins flipped by $\Lambda$. Thus, one can
construct scattering terms that are even/odd under time reversal by
taking
\begin{eqnarray}
U_{\pm} &=& U(x) [\cos(\Theta(\Lambda) - \alpha(x)) \nonumber \\
    &\pm& (-1)^Q\cos(\Theta(\Sigma_x\Lambda) +\alpha(x)) ]
\label{pert3}
\end{eqnarray}

Single
particle terms ($Q=1$) can arise from either a Zeeman interaction with a
magnetic field in the $xy$ plane, or a spin-orbit interaction. A
Zeeman interaction - which is odd under time reversal - generates
terms of the form $U_Z(x)\cos{(\Theta(\Lambda_{Z})+\alpha(x))}$,
with $Q(\Lambda_Z)=1$ and $\Sigma_x\Lambda_Z=-\Lambda_Z$. For large
$U_Z$ with appropriate spatial dependence, this term introduces a
mass term to $\Theta$. Including all the possibilities for $\Theta$,
such an interaction is sufficient to gap all the edge
modes. On the other hand, a spin-orbit coupling - which does not break
time reversal symmetry - generates two other kinds of terms. The
first type scatters an electron from an edge mode to its
time-reversed partner. It is of the form
$U_{so}(x)\partial_t\Theta \cos(\Theta + \alpha(x))$. This term cannot
gap any modes as it is a complete time derivative.
The second type is of the form $U_+$ of (\ref{pert3}), again with
$Q=1$. Such terms can gap out modes (but not
all of them, if the system is a topological insulator). Terms
that flip two spins can arise from time reversal symmetric
electron-electron interactions. Although not easily realized, similar
terms flipping more than two spins can also occur in principle.

We now examine whether a time reversal symmetric perturbation of
the type (\ref{pert3}) can gap the spectrum without spontaneously
breaking time reversal symmetry. We focus on the cases $N=1,2$ - the
analysis for larger $N$ is similar. In the single edge mode case,
$N=1$, we have $\sigma_{sH}/e^*=1$ for all $K$. Thus, according to the flux
insertion argument, no gap can be opened without breaking time
reversal symmetry in any
of these states. To see this microscopically, note that the only
charge conserving vectors are of the
form $\Lambda=(n,-n)$. The corresponding perturbation is even
under time reversal for even $n$ and odd for odd $n$. Thus,
perturbations of the form $U_+$ of (\ref{pert3}) require even $n$, say $n=2$.
For large $U$, such a perturbation can open a gap in the spectrum.
However, hand in hand with that it also spontaneously breaks time
reversal symmetry: when $U$ is large, the operator $\cos{\left
[\frac{e}{e^*}(\phi^\up(x)+\phi^\down(x))\right]}$, which
corresponds to $n=1$ and is odd under time
reversal, acquires a non-zero expectation value. Hence, it is
impossible to gap the two edge modes without breaking time
reversal symmetry, explicitly or spontaneously.

%Said differently, interaction between electrons on the two
%counter-propagating edge modes is of the form
%$\sum_{k,k',q,\sigma}V(q)c_{k+q,\sigma}^+
%c_{k,\sigma}c_{k'-q,-\sigma}^+c_{k',-\sigma}$. If the combination
%$\sum c_{k+q,\sigma}^+c_{k,-\sigma}$ assumes an expectation value (a
%situation that amounts to a ferromagnetic instability of a non-zero
%wave vector), the electron-electron interaction term becomes
%effectively a Zeeman term.

The case $N=2$ is a bit more complicated. In this case, the two pairs of
counter-propagating edge modes can be gapped if one can find two
linearly independent charge conserving $4$-component integer vectors
$\Lambda_1,\Lambda_2$ such that (a) $\Lambda_1=-\Sigma_x\Lambda_2$
and (b) $\Lambda_1^T \mathcal{K} \Lambda_1 = \Lambda_2^T \mathcal{K}
\Lambda_2 = \Lambda_1^T \mathcal{K} \Lambda_2 = 0$. Here, the second
condition is simply Haldane's criterion for gapping out
FQH edge modes. \cite{H9590} It guarantees that one can make a linear
change of variables from $\Phi$ to $\Phi'$ such that the action for
$\Phi'$ will be that of two decoupled non-chiral Luttinger
liquids. The two terms in (\ref{pert3}) will then gap the spectrum of
these two liquids by freezing the values of $\Theta(\Lambda_1)$ and
$\Theta(\Lambda_2)$.

In some cases, this freezing of these
values can lead to a spontaneous breaking of time reversal symmetry. As
in the $N=1$ case, this can happen if the perturbation fixes the
value of $\Theta(\Lambda)$ where $\Lambda$ is non-primitive - e.g. a
multiple of an integer valued vector. Then an operator of the form
$U_-$ of (\ref{pert3}) may acquire an expectation value.

We now turn to search for $\Lambda_1$ and $\Lambda_2$. It is convenient
to work in a basis where all four components of the charge
vector are $1$. We can parameterize the matrix $K$ as
\begin{equation}
K = \bpm b+us & b \\ b & b+vs \epm
\end{equation}
with $b,u,v,s$ integers and $u$ and $v$ having no common factor. In
terms of these parameters, the spin Hall conductance is
${(u+v)}/{\left [(u+v)b+uvs\right ]}$ and the elementary charge is
${1}/\left [{(u+v)b+uvs}\right ]$. Their ratio is then $(u+v)$, so
according to the flux insertion argument the parity of $u+v$
determines whether the spectrum can be gapped.

When $u+v$ is odd, it is indeed impossible to find $\Lambda_1,
\Lambda_2$ that satisfy (a), (b) and do not spontaneously break time
reversal symmetry. Imagine one had such
a solution and define $\Lambda_{\pm} = \Lambda_1 \pm \Lambda_2$. Then,
$\Sigma_x \Lambda_{-} = \Lambda_{-}$ and $Q(\Lambda_{-}) = 0$ so
$\Lambda^T_{-}$ must be an integer multiple of $(1,-1,1,-1)$. Also,
$\Sigma_x \Lambda_{+} = -\Lambda_{+}$ and $\Lambda_{-}^T \mathcal{K}
\Lambda_+ = 0$ so $\Lambda^T_+$ must be an integer multiple of
$(v,u,-v,-u)$. But $\cos(\Theta(v,u,-v,-u))$ is odd under time
reversal. This means that the scattering term corresponding to
$\Lambda_1, \Lambda_2$ will spontaneously break time reversal
symmetry.

On the other hand, when $u+v$ is even (so that both $u,v$ are
odd), this analysis suggests an obvious solution $(\Lambda_1,
\Lambda_2)$.  We can take $\Lambda^T_{-} = (1,-1,1,-1)$,
$\Lambda^T_{+} = (v,u,-v,-u)$, so that
\begin{equation}
\Lambda^T_1 = \frac{1}{2}\left(1+v, -1+u, 1-v, -1-u \right)
\label{evennumv}
\end{equation}
and $\Lambda_2 = -\Sigma_x \Lambda_1$. Note that
the scattering terms corresponding to $\Lambda_1,\Lambda_2$
flip the spins of $(v+u)/2$ electrons so at second order they flip
$(u+v)$ electrons - precisely the number needed to connect the two
degenerate states discussed in the flux insertion argument.

\textsl{Partner states:}
Now that we have derived the $\sigma_{sH}/e^*$ criterion using
two different approaches, we return to ask whether the
imposition of time reversal symmetry generally causes
each interacting universality class to split into two (or
more) subclasses, as it does in the non-interacting case.
Equivalently, does each fractional topological insulator have a
partner fractional trivial insulator (and vice versa) which
can be adiabatically connected to it in the absence of time reversal
symmetry?

The answer to this question depends on the parity of $1/e^*$ (at
least for the models analyzed here). When $1/e^*$ is odd,  a
partner state $\Psi'$ may be constructed for an arbitrary $s^z$
conserving fractional topological/trivial insulator $\Psi$. Let
$\Psi'$ be a decoupled bilayer state, where one layer is the
original state $\Psi$, and the other layer is a non-interacting,
$s^z$ conserving, topological insulator with spin-up/spin-down
electrons at $\nu = \pm k$ ($k$ odd). Clearly the spin-Hall
conductance of $\Psi'$ is given by $\sigma_{sH}' = \sigma_{sH} +
k$ where $\sigma_{sH}$ is the spin-Hall conductance of $\Psi$. Also, 
the elementary charge is the same as $\Psi$: $e'^*
= e^*$. Combining these two relations, we see that
$\sigma_{sH}'/e'^* = \sigma_{sH}/e^* + k/e^*$. Since $k, 1/e^*$
are odd, $\sigma_{sH}'/e'^*$ has the opposite parity from
$\sigma_{sH}/e^*$. Hence $\Psi'$ has a protected edge mode if and
only if $\Psi$ does not. On the other hand, it is clear that
$\Psi, \Psi'$ are adiabatically equivalent in the absence of time
reversal symmetry since this is true for non-interacting
topological/trivial insulators. We conclude that $\Psi'$ is indeed
a partner state to $\Psi$. Thus, the universality classes with
$1/e^*$ odd split into (at least) two subclasses when time
reversal symmetry is imposed.
% - just like the non-interacting case.

In contrast, we now show that for an even $1/e^*$ there are no
fractional topological insulators (at least for the models
analyzed here), since $\sigma_{sH}/e^*$ is necessarily even. 
For simplicity we prove the analogous statement
for quantum Hall states, e.g. $1/e^*$ even implies
$\sigma_{xy}/e^*$ even. Imagine adiabatically inserting $N =1/e^*$ 
flux quanta at a point $z_0$. As a consequence, a quasiparticle 
excitation will be created at $z_0$. We can compute the statistical 
angle for these excitations in two ways.
The first way is to explicitly compute the Berry phase associated
with exchanging two excitations. Using (\ref{berry}) with $N =
1/e^*$ and $q = \sigma_{xy}/e^*$ and dividing the result by $2$
since we are interested in an exchange rather than a $2\pi$
braiding, the Berry phase is
\begin{equation}
\theta = \pi \cdot \frac{1}{e^*} \cdot \frac{\sigma_{xy}}{e^*}
\end{equation}
Note that the coefficient of $\pi$ is a product of an integer
$\sigma_{xy}/e^*$ and an even integer $1/e^*$. The Berry
phase is thus a multiple of $2\pi$ - implying that the particles
are bosons. On the other hand,  this quasiparticle
excitation is made up of $\sigma_{xy}/e^*$ electrons. We conclude
that $\sigma_{xy}/e^*$ is even.

\textsl{Summary of results:} In this paper, we have shown that an 
$s^z$ conserving model is a fractional topological insulator if and 
only if $\sigma_{sH}/e^*$ is odd, where $\sigma_{sH}$ is the 
spin-Hall conductance (in units of $e/2\pi$) and
$e^*$ is the elementary charge (in units of $e$).

We thank the US-Israel BSF, the Minerva foundation, Microsoft
Station Q, NSF DMR-05-29399, and the Harvard Society of Fellows
for financial support.

\bibliography{frtopinsh}
\end{document}